\documentclass[useAMS,usenatbib]{mn2e}
\usepackage{graphicx,psfig,cite,epsf}  
\usepackage{times}
\usepackage{subfigure}
\usepackage{textcomp}
\usepackage{multirow}

\def\apjl{ApJ}
\def\apjl{ApJ}                
\def\apjs{ApJS}               
\def\ao{Appl.~Opt.}           
\def\aap{A\&A}                

\title[Broadband spectral analysis of SAX J1748.9-2021]{Broad-band spectral analysis of the accreting millisecond X-ray pulsar SAX J1748.9-2021}
\author[Pintore et al.] { Pintore F. $^{1,2}$\thanks{pintore@iasf-milano.inaf.it,\newline fabio.pintore@dsf.unica.it}, Sanna A.$^1$, Di Salvo T. $^3$, Del Santo M. $^4$, Riggio A.$^1$, D'A\`i A.$^4$, Burderi L.$^1$,\newauthor Scarano F.$^1$, Iaria R.$^3$ \\
$^1$ Universit\`a degli Studi di Cagliari, Dipartimento di Fisica, SP Monserrato-Sestu, KM 0.7, 09042 Monserrato, Italy\\
$^2$ INAF-Istituto di Astrofisica Spaziale e Fisica Cosmica - Milano, via E. Bassini 15, I-20133 Milano, Italy \\
$^3$ Dipartimento di Fisica e Chimica, Universit\`a di Palermo, via Archirafi 36 - 90123 Palermo, Italy \\
$^4$ INAF - Istituto di Astrofisica Spaziale e Fisica Cosmica - Palermo, Via U. La Malfa 153, I-90146 Palermo, Italy }

\begin{document}

\maketitle

\begin{abstract}

We analyzed a 115 ks {\it XMM-Newton} observation and the stacking of 8 days of {\it INTEGRAL} observations, taken during the raise of the 2015 outburst of the accreting millisecond X-ray pulsar SAX J1748.9-2021. 
The source showed numerous type-I burst episodes during the {\it XMM-Newton} observation, and for this reason we studied separately the persistent and burst epochs. { We described the persistent emission with a combination of two soft thermal components, a cold thermal Comptonization component ($\sim$2 keV) and an additional hard X-ray emission described by a power-law ($\Gamma\sim2.3$). 
The continuum components can be associated with an accretion disc, the neutron star (NS) surface and a thermal Comptonization emission coming out of an optically thick plasma region, while the origin of the high energy tail is still under debate. In addition, a number of broad ($\sigma=$ 0.1--0.4 keV) emission features likely associated to reflection processes have been observed in the {\it XMM-Newton} data. }
The estimated 1.0--50 keV unabsorbed luminosity of the source is $\sim$5$\times10^{37}$ erg s$^{-1}$, about 25$\%$ of the Eddington limit assuming a 1.4 M$_{\odot}$ NS. We suggest that the spectral properties of SAX J1748.9-2021 are consistent with a soft state, differently from many other accreting X-ray millisecond pulsars which are usually found in the hard state. 
Moreover, none of the observed type-I burst reached the Eddington luminosity. Assuming that the burst ignition and emission are produced above the whole NS surface, we estimate a neutron star radius of $\sim7-8$ km, consistent with previous results.

\end{abstract}

\begin{keywords}
accretion, accretion discs -- X-rays: binaries -- X-Rays: galaxies -- X-rays: individuals: SAX J1748.9-2021
\end{keywords}

\section{Introduction}
\label{intro}

After the discovery, in 1998, of SAX J1808.4-3658 \citep{wijnands98}, the first accreting neutron stars (NS) hosted in a low mass X-ray binary system (LMXB) and showing a spin pulsation of the order of the millisecond, the class of the accreting X-ray millisecond pulsars (AMXP) increased nowadays up to seventeen sources, including the transitional millisecond pulsare PSR J1023+0038 \citep{archibald15} and XSS J12270 \citep{papitto13b}. Their spectral and temporal properties have been widely investigated (see \citealt{patruno12,burderi13}, and reference therein, for recent reviews), proving that the matter in the accretion disc around the NS can exert a torque able to accelerate the NS spin period down to a few milliseconds (1.6 ms--5.5 ms the pulsation range of those currently detected). 
When the magnetic field of the NS is strong enough ({ estimated values are between $10^8-10^{9}$ G}), it is able to force at least part of the accreting matter towards the magnetic NS polar caps. The X-ray photons created by the release of energy above the NS surface is then observed in terms of pulsation. 
The millisecond pulsation frequency, reached by the transfer of angular momentum provided by the accreting matter onto the NS (see e.g. \citealt{bhattacharya91} for a review), can be then seen thanks to the offset angle between the rotation axis and the magnetic axis. Due to the magnetic field, the accretion disc can be truncated at the so-called magnetospheric radius, usually expressed in terms of a fraction of the Alfv\'en radius \citep{ghosh08}. Investigations on the values assumed by the truncation radius were carried out by studies of both timing and spectral properties \citep[e.g.][]{burderi98,papitto09,cackett10,patruno09b,wilkinson11}.
In particular, from a spectral point of view, an estimate on the inner disc radius can be provided by the analysis of broad emission lines often seen in the spectra of AMXPs. These are usually observed at the energies of the K-shell iron transitions (6.4--7.0 keV) and likely produced by reflection off of hard photons from the surface of the accretion disc. In particular, the broadening of these lines is commonly explained with relativistic effects associated to the fast rotation of the accretion disc in the proximity of the strong gravitational field of the compact object \citep[e.g.][]{cackett10,papitto09}. Therefore, from the shape of the emission feature, it is possible to trace the location of the radius of the accretion disc where the line is formed. In addition to these features, AMXP spectra are often characterized by a multicolour blackbody emitted from the accretion disc and a comptonized emission probably produced in the regions closer to the NS, as the boundary layer \citep[e.g.][]{papitto09}. In some cases, an additional soft blackbody component is detected and likely associated to the NS surface emission \citep[e.g.][]{poutanen06b}. 

For this sub-class of X-ray systems, generally the coherent NS pulsation is persistent during the source outbursts, { with the exception of three AMXPs (Aql X-1, HETE 1900.1-2455 and SAX J1748.9-2021) that showed intermittent pulsations whose origin is still uncertain (\citealt{kaaret06}, \citealt{casella08b}, \citealt{messenger15}; \citealt{galloway07}, \citealt{papitto13}; \citealt{gavriil07}, \citealt{altamirano08}).} In this work, we focus on the AMXP SAX J1748.9-2021, discovered by {\it Beppo-SAX} during the 1998 outburst, located in the globular cluster NGC 6440 \citep{intZand99} at a distance of $\sim8.5$ kpc and 0.6 kpc above the Galactic plane \citep{martins80,ortolani94,kuulkers03,valenti07}. Several outbursts were observed in 1998, 2001, 2005 and 2010 \citep{intZand99,intZand01, verbunt00,markwardt05,patruno09} and a probable outburst in 1971  \citep{markert75} was {\it a posteriori} associated with the source. During the quiescence of the source, \citet{intZand01} were also able to univocally detect the companion star and \citet{altamirano08} estimated its mass range (0.1--1 M$_{\odot}$). SAX J1748.9-2021 showed an intermittent pulsation at 442.361 Hz from which it was possible to infer an orbital period of $\sim$8.76 hr and a projected semi-major axis of $\sim0.4$ light seconds \citep{altamirano08, patruno09}. Finally, a number of type-I X-ray bursts were found with the \textit{Rossi X-ray Timing Explorer} (RXTE), which presented a blackbody emission of a few keV temperatures evolving during the thermonuclear flashes produced on the NS surface \citep{galloway08}.

In this work, we investigate the broad-band spectral properties of the source SAX J1748.9-2021 making use of a very high quality \textit{XMM-Newton} observation and the stacking of \textit{INTEGRAL} observations during the 2015 outburst. The source went in outburst on February, 17th and firstly detected by {\it INTEGRAL} \citep{kuulkers15,bozzo15}. Type-I X-ray bursts were also observed during later {\it INTEGRAL} observations \citep{bozzo15b}. The whole outburst lasted for about 60 days, and it was largely monitored by {\it Swift}/XRT telescope.
The spectral capabilities of \textit{XMM-Newton} provide a powerful insight for the investigation of the X-ray emission of this object, and also allow refined studies of the iron emission line and other discrete features in this source.

\section{Data Reduction}
\label{data_reduction}

We analyzed an \textit{XMM-Newton} observation taken on March, 4th 2015 (Obs.ID. 0748391301) obtained with a ToO at the beginning of the 2015 outburst, and the whole available 2015 \textit{INTEGRAL} dataset between March, 5th--13th. 
\begin{figure}
\center
\includegraphics[height=6.7cm,width=8.5cm]{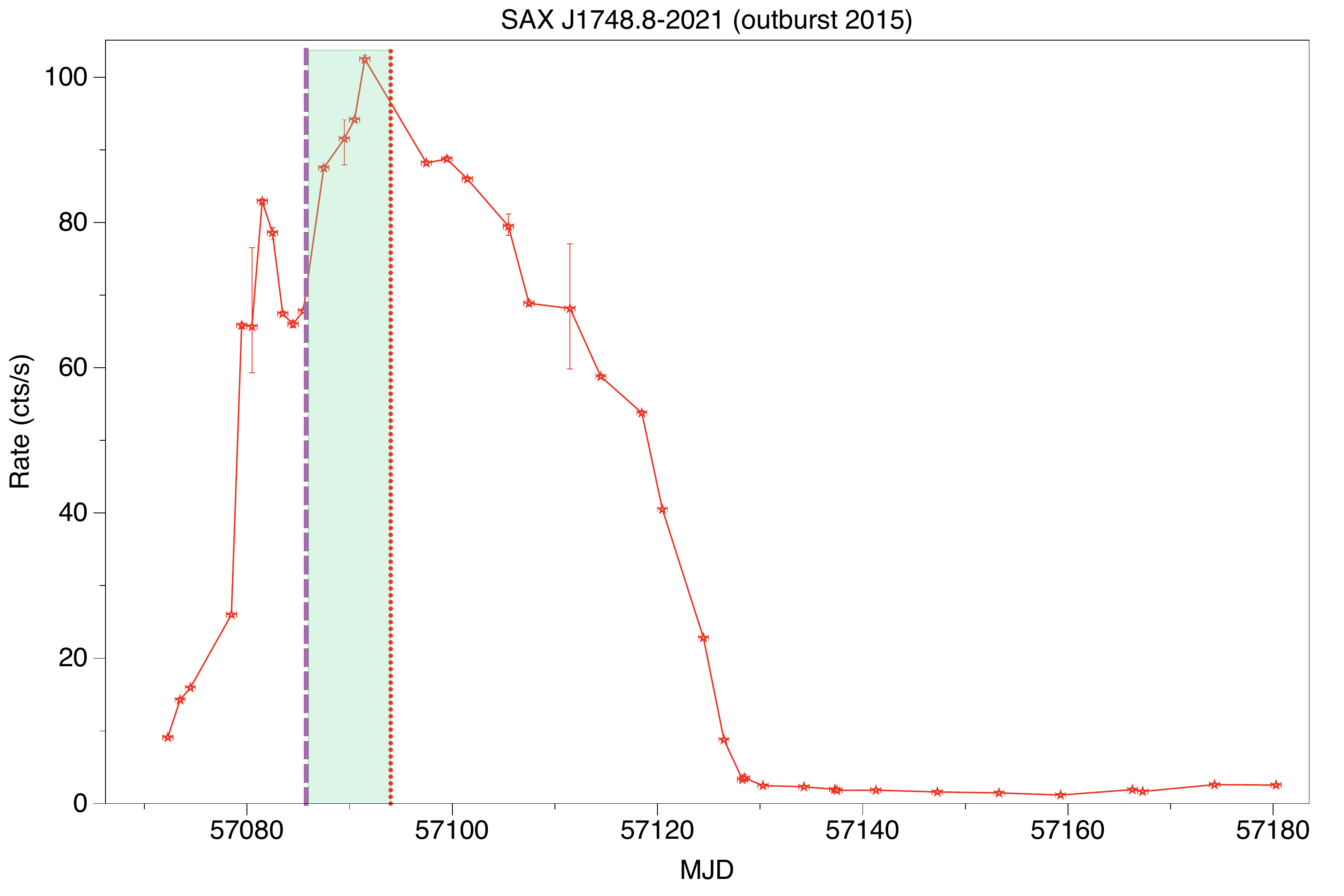}
\caption{{\it Swift}-XRT light curve of the 2015 outburst of SAX J1748.9-2021. The violet dashed line indicates the epoch of the {\it XMM-Newton} observation, the red dotted line is the epoch of the {\it INTEGRAL}/JEM-X observation, while the green area represents the observation epochs of {\it INTEGRAL}/ISGRI.}
\label{lc_swfit}
\end{figure}

The \textit{XMM-Newton} observation was made in both TIMING ($\sim$100 ks) and BURST ($\sim$10 ks) mode for the EPIC-pn. {  The reading times of the two modes are 30 $\mu$s and 7 $\mu$s, respectively, although for the BURST mode only the $3\%$ of the incident photons are collected}.
EPIC-MOS1 and EPIC-MOS2 were instead operating in TIMING and IMAGING mode, respectively. Data from all \textit{XMM-Newton} instruments were reduced using the up-to-date calibration files and the Science Analysis Software (SAS) v. 14.0.0 and adopting the standard reduction pipeline RDPHA (XMM-CAL-SRN-0312\footnote{http://xmm2.esac.esa.int/docs/documents/CAL-SRN-0312-1-4.pdf}; see also \citealt{pintore14b}). Due to severe pile-up in the EPIC-MOS data, which cannot be corrected by adopting the standard procedure of removing central columns (or pixels) of the CCD, we do not consider them in our analysis.
The last 5 ks of the TIMING mode EPIC-pn observation were affected by high solar background, hence we removed them from the analysis.
EPIC-pn spectra were extracted selecting events with {\sc pattern}$\leq 4$, which allows for single and double pixel events, and `{\sc flag}=0', which retains events optimally calibrated for spectral analysis. We remark that the EPIC-pn count rate observed during the TIMING mode is, on average, higher than 700 cts s$^{-1}$, and hence the {\it counting mode} was often triggered especially during the type-I bursts, limiting our studies on them.
Because of the high count rate of the source, pile-up effects are not negligible. In order to investigate pile-up effects in the data, we compared the spectrum extracted taking into account all the columns of the CCD in the range RAWX=[31:43] with spectra obtained excising one, three and five central brightest columns of the aforementioned RAWX range. We found that pile-up effects can be strongly limited by excising only the central column RAWX=37, therefore hereafter we will refer to the spectrum extracted in the range RAWX=[31:43] without the central column. The background spectrum was instead extracted in the RAWX range [3:5]. We checked that the background extracted in this region was not heavily contaminated by the source, comparing it with the local background estimated during an {\it XMM-Newton} observation when the source was in quiescence. The response matrix and the ancillary file were created following the standard procedure for the pile-up correction. 
The EPIC-pn spectrum was finally rebinned with an oversample of 3 channels per energy resolution element using the \textit{specgroup} task. 

The reduction of RGS spectra, first and second order, was carried out adopting the standard \textit{rgsproc} task, filtering for high background intervals. Pile-up effects are not negligible because the RGS1 and RGS2 fluxes indicate that the nominal pile-up threshold is overcame. In more detail, the inferred flux in the 0.3--2.0 keV (see Section~\ref{rgs}) for both instruments is $\sim$5.2$\times10^{-10}$ and $\sim$4.9$\times10^{-10}$ erg cm$^{-2}$ s$^{-1}$ (adopting a simple power-law model) which is well above the pile-up threshold of RGS1 and RGS2 reported by the {\it XMM-Newton} user manual ($\sim$3$\times10^{-10}$ erg cm$^{-2}$ s$^{-1}$ and $\sim$1.5$\times10^{-10}$ erg cm$^{-2}$ s$^{-1}$, respectively). Because pile-up in RGS cannot be easily solved, we will analyze RGS data separately from the EPIC-pn and {\it INTEGRAL} spectra.
We create an average spectrum stacking together the RGS1 and RGS2 data and, adopting the tool \textit{rgscombine} which combines, separately, first and second order spectra, we rebin the data with at least 1000 counts per bin. Because the background is dominant below 0.75 keV, we analyze the energy range 0.75--2.0 keV.

We searched for all the available {\it INTEGRAL} observations performed in 2015 between March, 5th and 13th.
In order to maximise the IBIS/ISGRI \citep{lebrun03} and JEM-X \citep[][]{lund03} spectral responses,
we selected only pointings (i.e. science windows, SCW, with typical duration of 2 ks)
with the source located within 7.5{\textdegree} and 3.5{\textdegree} from the centre of the IBIS and JEM-X field of views (FOV), respectively.
{Since the single observations do not show significant spectral variability amongst them (apart for flux changes)}, we stacked 16 SCWs for IBIS/ISGRI and 3 for JEMX-1 and JEMX-2 which have been analyzed
by using version 10.1 of the OSA software distributed by the ISDC \citep{courvoisier03}.
Because of the faintness of the source in hard X-rays (about 10 mCrab, or $\sim8\times10^{-11}$ erg cm$^{-2}$ s$^{-1}$, in 20--40 keV),
the ISGRI spectrum has been extracted by the total mosaic image with the {\sc mosaic$\_$spec} tool in 3 energy bins spanning from 20 keV up to 60 keV.
However, the source is very bright (about 100 mCrab) in the soft X-rays.
The source spectrum could be extracted from mosaic images of the two cameras of the JEM-X telescope,
i.e. JEMX-1 and JEMX-2.
First we produced images for each SCW in 16 narrow energy bands (from 3 to 35 keV) and then we extracted the spectrum by the total image with the {\sc mosaic$\_$spec tool}. However, according to the {\it INTEGRAL} calibration team,  spectra are actually well calibrated in the range 7-25 keV, thus the final JEM-X spectra have only 7 bins. The total good time intervals of the IBIS/ISGRI and JEM-X spectra are 30.5 ks and 5.1 ks, respectively.

We fit simultaneously the \textit{XMM-Newton} (EPIC-pn) and \textit{INTEGRAL} (JEMX-1, JEMX-2 and ISGRI) spectra, using {\sc xspec} v. 12.8.2 \citep{arnaud96}. We select the energy ranges 1.0--10.0 keV, 7.0--25.0 keV and 20.0--50.0 keV, for EPIC-pn, JEMX-1/X2 and ISGRI spectra, respectively. We ignore the EPIC-pn channels at energies lower than 1.0 keV as calibrations in TIMING mode are still uncertain below this energy (internal \textit{XMM-Newton} report CAL-TN-
0083\footnote{http://xmm2.esac.esa.int/docs/documents/CAL-TN-0083.pdf}). We also consider a systematic error of 0.15$\%$ and and 2$\%$ for the EPIC-pn and {\it INTEGRAL} spectra, respectively.

\begin{figure}
\center
\includegraphics[height=6.7cm,width=8.5cm]{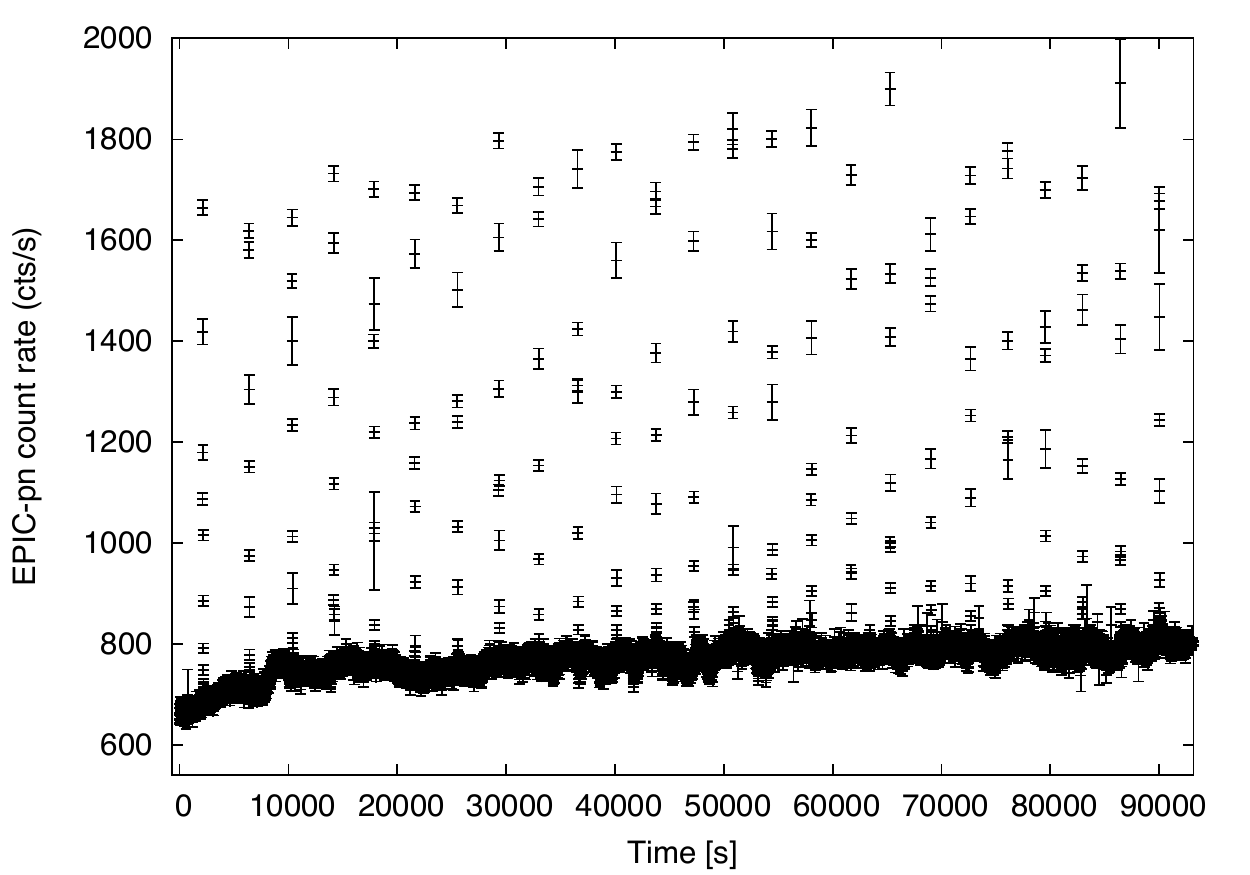}
\caption{{\it XMM-Newton} EPIC-pn 0.3-10 keV light curve of SAX J1748.9-2021 after the removal of the latest 10ks of data affected by high solar background. 25 type-I X-ray bursts were detected during the observation.} 
\label{lc_tot}
\end{figure}

\begin{figure}
\center
\includegraphics[height=6.7cm,width=8.5cm]{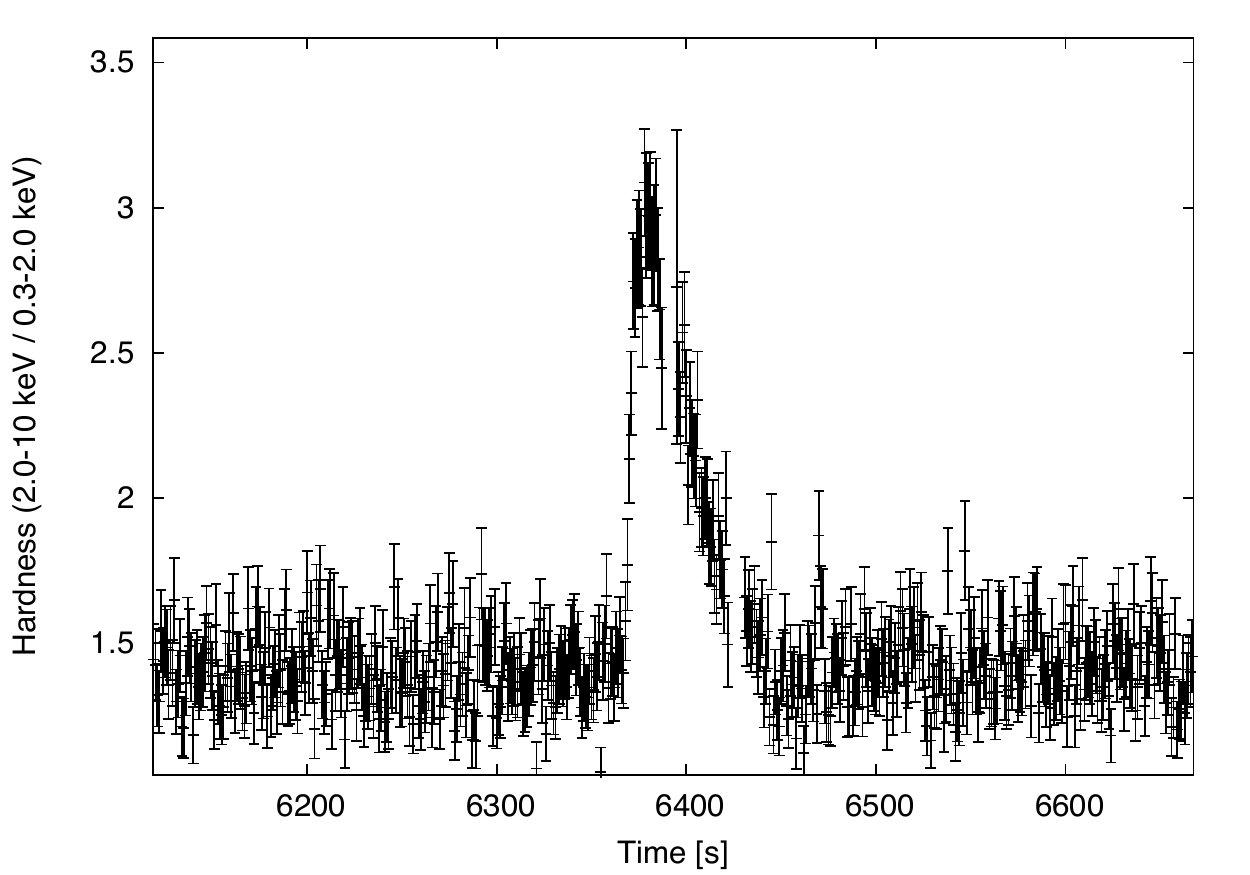}
\caption{Hardness ratio of the 2.0--10.0 keV and 0.3--2.0 keV {\it XMM-Newton} EPIC-pn light curves of SAX J1748.9-2021, around a type-I burst. Time is calculated from the beginning of the {\it XMM-Newton} observation. The hardness ratio is quite constant during the persistent emission, but clear variability is introduced during the burst episode.} 
\label{hard_burst}
\end{figure}

\section{Spectral analysis}
\label{analysis}

The \textit{XMM-Newton} observation caught SAX J1748.9-2021 during the initial phases of the outburst, approximately at half of the peak luminosity. The peak was instead caught by the \textit{INTEGRAL}-ISGRI and \textit{INTEGRAL}-JEMX observations (see Figure~\ref{lc_swfit}). In the \textit{XMM-Newton} observation, SAX J1748.9-2021 showed 25 type-I bursts during the EPIC-pn TIMING mode (see Figure~\ref{lc_tot}) and 2 during the EPIC-pn BURST mode. Since X-ray bursts affect the spectral properties of the persistent emission of the source (see the hardness ratios variability showed in Figure~\ref{hard_burst}), we firstly focus our spectral analysis on the persistent emission (i.e. without bursts), in Section~\ref{persistent}, and subsequently we investigate the properties of the type-I bursts in Section~\ref{burst_sec}. We note that the persistent emission does not show short-term spectral variability during the observation, hence, hereafter, we will refer only to the average persistent spectrum. In addition, because of the very poor counting statistics of the EPIC-pn BURST mode, we limit our analysis of the persistent epochs to the EPIC-pn TIMING mode only and we use both modes for the type-I burst analysis. We finally note that no bursts are found during the {\it INTEGRAL}/JEM-X considered SCWs { and that no spectral variability was found between EPIC-pn and {\it INTEGRAL} spectra, as the best-fit spectral model for the broadband spectra (see Section~\ref{persistent}) can also well fit the EPIC-pn spectrum alone.}

\subsection{Persistent emission}
\label{persistent}

\begin{figure}
\center
\includegraphics[height=9.2cm,width=7.5cm,angle=270]{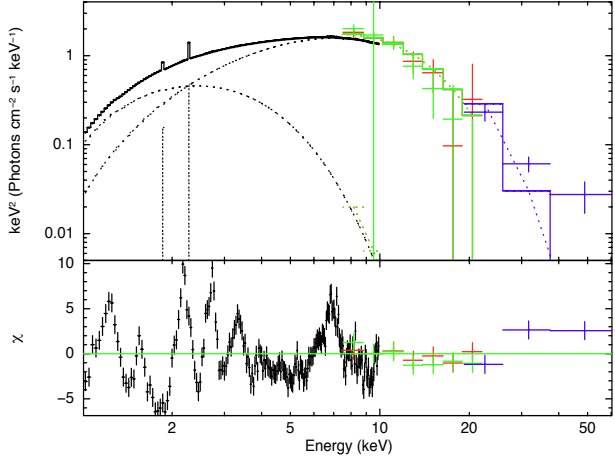}
\caption{Unfolded $E^2f(E)$ \textit{XMM-Newton}/EPIC-pn (\textit{black}), \textit{INTEGRAL}/JEMX-1 (\textit{red}) and JEMX-2 (\textit{green}), and \textit{INTEGRAL}/ISGRI (\textit{blue}) fitted with absorbed {\sc diskbb} and {\sc nthcomp} components. Several features in the EPIC-pn spectrum are observed across the 0.3--10 keV; an excess is also found above 20 keV in the ISGRI spectrum (see text).}
\label{continuum_resid}
\end{figure}

We initially fit EPIC-pn and {\it INTEGRAL} data together with a simple absorbed ({\sc tbabs} in {\sc xspec}) multicolour blackbody disc ({\sc diskbb}, \citealt{mitsuda84}) plus a thermal Comptonization model ({\sc nthcomp}, \citealt{zdiarski96,zycki99}). We adopt the abundances of \citet{anders89} and the cross-sections of \citet{balucinska92}. { For each spectrum, we add a multiplicative constant which takes into account the different cross calibrations of the instruments.} In addition, we note that the disc temperature and the seed photons temperature for Comptonization are very similar, hence we keep them tied (i.e. assuming that the soft component is the source of seed photons). We clearly identify structures in the EPIC-pn residuals, likely associated with emission features. Because the EPIC-pn spectra taken in TIMING mode are affected by systematic emission features at 1.8--1.9 keV and 2.2--2.3 keV produced by instrumental Si and Au edges, hereafter we will include two narrow gaussian emission lines in the model. 
{ We find that the fit with this model is statistically poor and leaves several residuals (Figure~\ref{continuum_resid}) associated with broad emission features and with an excess in the ISGRI data, meaning that at least an additional component is requested for the continuum. }

\begin{table}
\footnotesize
\begin{center}
\caption{Best fit spectral parameters obtained with the absorbed continuum {\sc bbody+diskbb+nthcomp+power-law} model, plus an absorption edge and 6 emission lines. Errors are at 90$\%$ for each parameter.}
\label{table_continuum}
\begin{tabular}{lll}
\begin{tabular}{lll}
\hline
Model & Component &\multicolumn{1}{c}{(1)} \\
\\
{\sc TBabs} & nH($10^{22}$$^a$ cm$^{-2}$) & $0.56^{+0.05}_{-0.03}$ \\
\\
{\sc diskbb} & kT$_{disk}$ (keV)$^b$ & $0.95^{+0.5}_{-0.4}$ \\
 & norm$^c$ & $108^{+54}_{-15}$ \\
\\
{\sc bbody} & kT$_{bb}$ (keV)$^d$ & $1.08^{+0.2}_{-0.06}$ \\
& Radius (km)$^e$ & $7 \pm 3$ \\
\\
{\sc nthComp} & $\Gamma_{compt}$$^f$ & $<1.07$ \\
 & kT$_e$ (keV)$^g$ & $2.10^{+0.1}_{-0.08}$ \\
 & norm(10$^{-3}$)$^h$ & $9^{+40}_{-2}$ \\
\\
{\sc power-law} & $\Gamma$$^i$ & $2.3\pm0.2$ \\
 & norm$^j$ & $0.066^{+0.068}_{-0.065}$ \\
\\
{\sc edge} & Energy (keV)$^k$ & $8.7\pm0.1$ \\
 & $\tau$$^l$ & $0.022\pm0.006$ \\
\\
{\sc diskline} & Energy (keV)$^m$ & $6.79^{+0.05}_{-0.06}$ \\
 & $\beta$$^n$ & $-2.7^{+0.1}_{-0.2}$ \\
 & R$_{in}$ (R$_{g}$)$^o$ & $29^{+12}_{-9}$ \\
 & R$_{out}$ (R$_{g}$)$^p$ & 10$^{5}$ (frozen) \\
 & Inclination (deg)$^q$ & $44^{+10}_{-6}$ \\
 & norm (10$^{-4}$)$^r$ & $9.9^{+1.6}_{-1.3}$ \\
\\
{\sc gaussian} & Energy (keV)$^s$ & $1.23\pm0.02$ \\
 & $\sigma$ (keV)$^t$ & $0.06^{+0.03}_{-0.3}$ \\
 & norm (10$^{-3}$)$^u$ & $1.5^{+0.8}_{-0.5}$ \\
\\
{\sc gaussian} & Energy (keV)$^s$ & $1.56\pm0.01$ \\
 & norm (10$^{-4}$)$^u$ & $5.9^{+1.4}_{-1.3}$ \\
\\
{\sc gaussian} & Energy (keV)$^s$ & $2.187^{+0.008}_{-0.009}$ \\
 & $\sigma$ (keV)$^t$ & $0.11\pm0.01$ \\
 & norm(10$^{-3}$)$^u$ & $2.4\pm0.3$ \\
\\
{\sc gaussian} & Energy (keV)$^s$ & $2.698\pm0.009$ \\
 & $\sigma$ (keV)$^t$ & $0.096^{+0.013}_{-0.006}$ \\
 & norm (10$^{-3}$)$^u$& $1.7^{+0.2}_{-0.1}$ \\
\\
{\sc gaussian} & Energy (keV)$^s$ & $3.29\pm0.02$ \\
 & $\sigma$ (keV)$^t$ & $0.17^{+0.03}_{-0.02}$ \\
 & norm (10$^{-3}$)$^u$ & $1.2\pm0.3$ \\
\\
{\sc gaussian} & Energy (keV)$^s$ & $3.98\pm0.07$ \\
 & $\sigma$ (keV)$^t$ & $0.1^{+0.1}_{0}$ \\
 & norm (10$^{-4}$)$^u$ & $1.9^{+1.5}_{-1.1}$ \\
\hline
		 	    &$\chi^2/dof$				   &199.53/183						\\
\end{tabular}
\end{tabular} 
\end{center}
\begin{flushleft} $^a$ Neutral column density; $^b$ multicolour accretion disc temperature; $^c$ normalization of the {\sc diskbb}; $^d$ blackbody temperature; $^e$ emission radius of the {\sc blackbody}; $^f$ photon index; $^g$ electrons temperature of the corona; $^h$ normalization of {\sc nthcomp}; $^i$ photon index of the {\sc power-law}; $^j$ normalization of the {\sc power-law}; $^k$ energy of the absorption edge; $^l$ optical depth of the edge; $^m$ energy of the {\sc diskline} emission line; $^n$ power law dependence of emissivity; $^o$ inner radius in units of gravitational radii R$_g$; $^p$ outer radius in units of gravitational radii R$_g$; $^q$ inclination angle of the binary system; $^r$ normalization of the {\sc diskline}; $^s$ energy of the emission line; $^t$ broadening of the line; $^u$ normalization of the line; \\
\end{flushleft}
\end{table}

{ We initially model the broad iron emission feature, found at $\sim 6.75$ keV. Because of its broadness ($\sigma\sim$0.4 keV), we fit it alternatively with a {\sc gaussian} model or with a relativistically smeared reflection profile ({\sc diskline}, \citealt{fabian89}), the latter in order to take into account probable effects due the fast rotation of the matter in the disc if the line was produced by reflection off of hard photons from the surface of the inner disc. We find that, although the introduction of a {\sc gaussian} model is statistically significant, the {\sc diskline} provides a marginally better fit to the data ($\Delta\chi^2=220.89$ for 3 additional dof versus $\Delta\chi^2=242.44$ for 5 additional dof, respectively). { Moreover, the {\sc diskline} model allows us to find a coherent picture with the disc model (see below).}
For this reason, in the following, we adopt the {\sc diskline} model for the iron line. Because the fit is insensitive to the parameter, we fix the outer disc radius to $10^5$ gravitational radii (R$_g=GM/c^2$ where G is the gravitational constant, M is the NS mass and c is the speed of light). The energy of the iron emission feature ($\sim6.7/6.8$ keV) is  consistent with a Fe XXV K$_{\alpha}$ line (EW of $\sim$30 eV), produced in the disc at a distance of $\sim20-43$ R$_g$. In addition, from the line, we can also constrain the inclination angle of the system being in the range 38\textdegree-- 45\textdegree. 
On the other hand, we note that the gravitational radius at which the reflection line should have been produced (20--43 R$_g$) is not consistent with the inner disc temperature ($\sim1.2\pm0.1$ keV), which instead suggests an inner disc radius of $(6-8)/cos(\theta)$ km (where $\theta$ is the inclination angle, and it cannot be likely higher than 70--75{\textdegree} as no eclipses are seen for this source), thus apparently below the latest stable orbit and physically not realistic. 

The combination of a small inner disc radius, a high temperature and the discrepancy with the {\sc diskline} inner radius makes the fit with a single soft component unlike. A more physical spectral description suggests instead that the soft component may be split in two components, a multicolour blackbody ({\sc diskbb}) to model the emission from the accretion
disc, and a blackbody to model the emission from the NS surface. For this model, we keep linked the blackbody temperature and the seed photons temperature of the {\sc nthcomp}. We find an absorption neutral column density of nH$\sim5.8\times10^{21}$ cm$^{-2}$ (about a factor of 2 higher than the average Galactic extinction in the direction of the source, $3.29\times10^{21}$ cm$^{-2}$; \citealt{dickey90}), an inner disc temperature of kT$_{disc}\sim0.95$ keV, a NS surface temperature of kT$_{disc}\sim1.1$ keV, and an electron temperature of the comptonizing plasma of kT$_{e}\sim2.1$ keV with an asymptotic power-law photon index $<1.1$. These spectral parameters suggest that the source was observed during a soft state. 
The radius related to the emission of the {\sc bbody} component is consistent with either the NS surface or a region very close to the NS (R$=7\pm3$ km).
From the {\sc diskbb} temperature (which, taking into account the uncertainties, can be as low as 0.6 keV) and normalization is possible to determine an upper value to the inner disc radius of R$_{in}\sim15$ km (for an inclination of 37--56{\textdegree}, 90$\%$ error, and not corrected for any color temperature factor) which could be compared to the lower limit of the inner disc radius inferred by the {\sc diskline} model ($\sim38$ km). 

However, residuals are still observed at very high energy in the ISGRI spectrum. Since hard power-law tails are often found in the spectra of Galactic black hole \citep[e.g.][]{mcclintock06,delsanto15} and neutron star X-ray binary systems \citep[e.g.][]{disalvo00,disalvo01,iaria04,disalvo06,paizis06,dai07,tarana07,piraino07,delsanto08,bouchet09,delsanto12}, we add a {\sc power-law} (in {\sc xspec}) component to the total model. This component is able to flatten the higher energy residuals and it is statistically significant ($\Delta\chi^2 = 17$ for 2  additional dof). {This finding indicates that the thermal Comptonization component alone is not able to describe the hard X-rays emission while a further non-thermal component is necessary. The physical meaning of additional hard X-ray power-laws in spectra of XRBs is still matter of debate. They have been associated with either emission of Comptonization by a non-thermal medium \citep{poutanen98} or bulk motion of accreting material close to the NS (e.g. \citealt{titarchuk98}, but see also \citealt{disalvo00}).} {We verified that such a feature is not due to the predominance of the EPIC-pn spectrum guiding the fit: by excluding the EPIC-pn spectrum and fitting simultaneously only the ISGRI, JEMX-1 and JEMX-2 data, we can still find evidence of this hard tail. Moreover, we tried with a {\sc compPS} model \citep{poutanen96} fitting the {\it INTEGRAL} data assuming both the Maxwellian electron distribution and the hybrid (thermal/non-thermal) distribution. We found that the fit is statistically acceptable only when using the hybrid model. However, because of the low statistics at high energy, we did not manage to constrain the electron power-law index.}

{The interpretation of the continuum provides a coherent picture of the accretion mechanisms of the source. In fact, millisecond coherent pulsations at $\sim$442 Hz have been found in the {\it XMM-Newton} observation (Sanna et al. submitted), pointing towards a magnetically driven accretion column onto the NS polar caps, which produces most of the blackbody emission. In addition, this scenario rules out the possibility that the disc reached the last stable orbit, where the magnetic lines cannot channel the matter anymore towards the polar caps, but instead supports a truncated accretion disc, as that inferred in the best-fit continuum model.}}

\begin{figure}
\center
\includegraphics[height=9.2cm,width=7.5cm,angle=270]{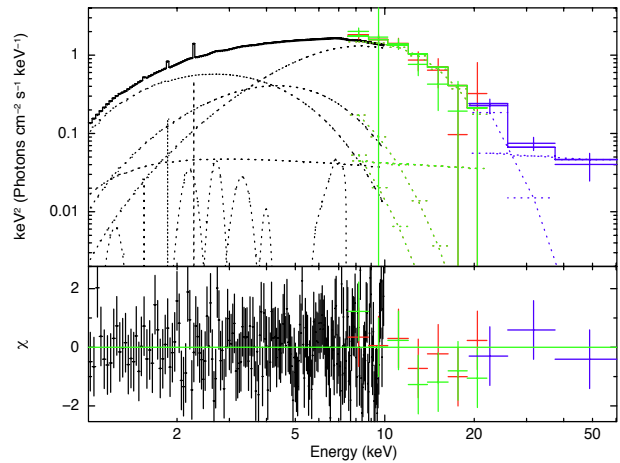}
\caption{Unfolded $E^2f(E)$ \textit{XMM-Newton}/EPIC-pn (\textit{black}), \textit{INTEGRAL}/JEMX-1 (\textit{red}) and JEMX-2 (\textit{green}) and \textit{INTEGRAL}/ISGRI (\textit{blue}). The dashed curves represent the best-fit continuum ({\sc diskbb}, {\sc bb}, {\sc nthcomp} and {\sc power-law} components) plus 9 additional narrow and broad emission lines (see text).}
\label{continuum}
\end{figure}

In addition to the continuum, we also mention that emission features with centroid energy of 1.23 keV, 1.56 keV, 2.18 keV, 2.7 keV, 3.29 keV and 3.98 keV are found; however the lines at 1.56 keV and 2.18 keV are likely dubbed instrumental as the former (probably Al) has a broadening consistent with 0, while the latter is not easily associated with any of the known X-ray transitions of the most abundant elements. In particular, the 2.18 keV line is close to the Au instrumental edge of the EPIC-pn, and therefore it may be due to a miscalibration of the response matrix in the $2.2-2.3$ keV energy range. The other lines may be instead associated to K-shell emission lines of Na IX (1.237 keV), S XVI (2.62 keV), Ar XVIII (3.32 keV) and Ca XX or Ca XIX (4.11 keV or 3.90 keV). However, we are suspicious about the Na IX line as the chemical abundance of this element is low if compared to the ionized transitions of other more abundant elements which can show emission line at energies $<2$ keV (for example Mg XI, H-like and He-like lines at $\sim$1.47 keV and $\sim$1.35 keV, respectively). On the other hand, we note that the neutral Mg has a K-$\alpha$ transition at 1.25 keV: if the observed line is associated with neutral Mg, we suggest that such a line is produced in a different region than the other emission features. Hereafter we limit our discussion to the lines of S, Ar and Ca: their broadening is usually lower than 0.2 keV (see Table~\ref{table_continuum}) and their equivalent width (EW) is consistent with $\sim$10 eV, $\sim$10 eV and $\sim$2 eV, respectively. 
{ Because the broadening of the Fe, Ar, Ca and S lines is similar to that of the iron line, we tentatively describe also the low energy lines with {\sc diskline} models, linking all the relativistic parameters to those of the iron {\sc diskline} component \citep[as in][]{disalvo09a}. However, this approach provides a poorer fit with respect to {\sc gaussian} models applied to the low energy lines, $\chi^2_{\nu}$=1.22 versus $\chi^2_{\nu}$=1.1; in particular, with respect to the continuum model without the low-energy emission lines, the F-test probability of chance improvement is $<3\times10^{-37}$ and $<4.9\times10^{-40}$ when adopting {\sc diskline} and {\sc gaussian} models, respectively. Furthermore, we note that, directly comparing the {\sc gaussian} and {\sc diskline} models, the former is statistically better with an F-test probability of chance improvement $<2\times10^{-5}$.
Hence, hereafter, we will only discuss the fit with the low energy lines fitted with {\sc gaussian} models}.
The final fit is clearly robust and well describes the broadband spectrum ($\chi^2$/dof=199.53/183). We report the best fit parameters in Table~\ref{table_continuum} and show the best fit spectrum in Figure~\ref{continuum}. 

If the large number of emission features is possibly associated to hard photons reflected from the disc, it suggests to fit the broadband spectrum with a self-consistent reflection model. We test the {\sc reflionx} \citep{ross05} and the {\sc rfxconv} \citep[e.g.][]{kolehmainen11} models. We remove the Fe {\sc diskline} model and we left only the gaussian models for those low energy lines not included in the models (specifically Ar and Ca lines). { Here we discuss only the {\sc rfxconv} model for sake of simplicity as the {\sc reflionx} model gives very similar results. We convolve the {\sc rfxconv} with the {\sc rdblur} model in order to take into account the relativistic effects. We find that the best-fit model does not provide any statistical improvement (reduced $\chi^2$ of 1.2) with respect to the fit with {\sc diskline} and {\sc gaussian} models. Furthermore, the parameters of the {\sc rdblur} are highly unconstrained. However the self-consistent model allows us to find that the reflection fraction is $0.094\pm0.008$, produced in a medium with an ionization of Log$\xi$$\sim$3.2. We also tentatively let the iron abundance to vary but the fit resulted insensitive to any value assumed by the parameter. Hence, we cannot determine a stable continuum model for reflection and we should limit our discussion to the reflection fraction and the ionization parameter of the disc.
Finally, we remark that, also with the self-consistent models, the hard {\sc power-law} component is still requested by the spectral fit with an F-test probability of chance improvement $< 10^{-5}$.}

\subsection{RGS}

We avoid to fit the persistent RGS data (i.e. without bursts) together with the EPIC-pn spectrum using the same spectral model because of a clear discrepancy between RGS and EPIC-pn data. Although this is not dramatic, it introduces significant residuals in the simultaneous fit with the EPIC-pn data. We suspect this is not produced by a miscalibration of the EPIC-pn energy scale but rather from pile-up effects in the RGS and some other residuals miscalibration of these detectors. Indeed, following the standard procedure suggested in the {\it XMM-Newton} thread\footnote{http://xmm.esac.esa.int/sas/current/documentation/threads/Pile-Up\_in\_the\_RGS.shtml}, we infer that pile-up effects are present in these data.

\begin{table}
\footnotesize
\begin{center}
\caption{RGS data, best fit spectral parameters obtained with the absorbed continuum {\sc diskbb+nthcomp+power-law} model, plus three absorption edge and 6 absorption lines. Errors are at 90$\%$ for each parameter.}
\label{RGS_fit}
\begin{tabular}{lll}
\hline
Model & Component &\multicolumn{1}{c}{(1)} \\
\\
{\sc tbabs} 	    &N$_H$ (10$^{22}$ cm$^{-2}$)$^a$ &$0.572^{+0.002}_{-0.002}$ 	 \\
\\
{\sc diskbb} 	    &kT$_{bb}$ (keV)$^b$			    & $2.95^{+0.08}_{-0.3}$	  \\
& Norm. $^c$ 						    		    & $5.59^{+0.01}_{-0.02}$ 		 \\
\\
{\sc edge} 	     & Energy (keV)$^d$			    & 0.8691$_{-0.007}^{+0.008}$ 		\\
& {$\tau_{edge}$} $^e$ 						    & 0.358$_{-0.008}^{+0.009}$ 		  \\
\\
{\sc edge} 	     & Energy (keV)$^d$			    & 1.313$_{-0.002}^{+0.003}$ 		\\
& {$\tau_{edge}$} $^e$ 						    & 0.101$_{-0.007}^{+0.004}$ 		  \\
\\
{\sc edge} 	     & Energy (keV)$^d$			    & 1.523$_{-0.006}^{+0.004}$ 		\\
& {$\tau_{edge}$} $^e$ 						    & 0.096$_{-0.005}^{+0.004}$ 		  \\
\\
{\sc gaussian} 	     & Energy (keV)$^f$			    & 0.923$\pm0.02$ 		\\
& Norm. ($10^{-4}$)$^g$ 						    & 6$\pm2$				  \\
\\
{\sc gaussian} 	     & Energy (keV)$^f$			    & 0.963$\pm0.001$ 		 \\
& Norm. ($10^{-4}$)$^g$ 						    & 4.4$\pm2$				  \\
\\
{\sc gaussian} 	     & Energy (keV)$^f$			    & 1.121$_{-0.002}^{+0.004}$ 		 		 \\
& Norm. ($10^{-4}$)$^g$ 						    & 4$\pm1$				  \\
\\
{\sc gaussian} 	     & Energy (keV)$^f$			    & 1.177$_{-0.005}^{+0.008}$ 			 \\
& Norm. ($10^{-4}$)$^g$ 						    & 3$\pm1$				  \\
\\
{\sc gaussian} 	     & Energy (keV)$^f$			    & 1.381$_{-0.009}^{+0.01}$ 		 			\\
& Norm. ($10^{-4}$)$^g$ 						    & 2$\pm1$	  				\\
\\
{\sc gaussian} 	     & Energy (keV)$^f$			    & 1.7680$_{-0.004}^{+0.0009}$ 	 		\\
& Norm. ($10^{-3}$)$^g$ 						    & 1.3$\pm0.2$ 	  				\\
\\
\hline
		 	    &$\chi^2/dof$				   &1178.26/631					\\
\end{tabular} 
\end{center}
\begin{flushleft} $^a$ Neutral column density; $^b$ multicolour accretion disc temperature; $^c$ normalization of the {\sc diskbb}; $^d$ energy of the absorption edge; $^e$ optical depth of the edge; $^f$ energy of the emission line; $^g$ normalization of the line. \\
\end{flushleft}
\end{table}

\begin{figure}
\center
\includegraphics[height=9.2cm,width=7.5cm,angle=270]{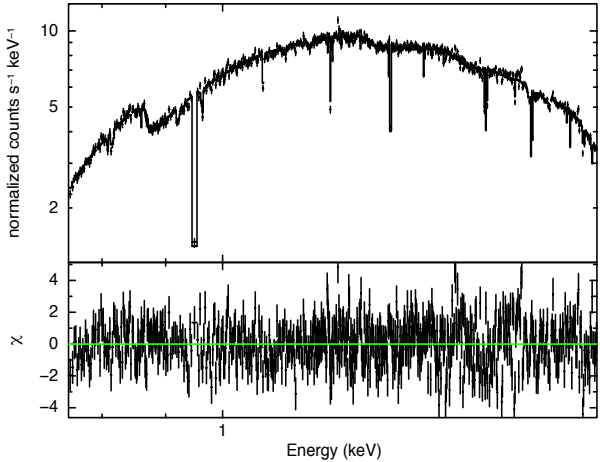}
\caption{Unfolded $E^2f(E)$ \textit{XMM-Newton}/RGS. A number of absorption features were added to the continuum and some edge were left free to vary (see text).}
\label{rgs}
\end{figure}

We initially fit the stacked spectrum with an absorbed {\sc diskbb} model and we obtain a very poor fit ($\chi^2/dof=2468.71/650$; nH$\sim0.57 \times 10^{22}$ cm$^{-2}$ and kT$_{disc}\sim2$ keV). Large residuals are found in the fit, in particular around the neutral edges of Ne, Mg and Al. We may explain them as a mismatch in the depths associated with their neutral edges, probably due to the data being sensible to differences in the abundances of the different elements with respect to the adopted solar abundance table \citep{anders89}. Therefore, we substitute the {\sc tbabs} model, with the {\sc tbvarabs} model, where it is possible to independently vary the abundance value for each element. We leave free to vary the abundance of Ne, Mg and Al and we found an improvement in the fit ($\chi^2/dof=1557.14/647$); the abundances of Ne and Mg converged towards $\sim$1.8 and $\sim$2.5 times the solar abundance, while the Al abundance was unlikely high ($>10$ times the solar abundance). The latter result suggests that the Al edge might not have an astrophysical nature and we very tentatively suggest instrumental effects, also related to the narrow Al emission line found in the EPIC-pn spectrum. 
In any case, residuals are still observed around the energies of these edges. To account for this issue, in {\sc tbvarabs} we fix at zero the abundances of Ne, Mg and Al and we add three absorption edges to the model, leaving their energy and optical depth free to vary. This model further improves the fit ($\chi^2/dof=1383.28/644$), showing that the energies of Ne and Mg are slightly different but consistent within the uncertainties with those obtained in laboratory (Table~\ref{RGS_fit}). We note that the Ne edge can be also confused with the O VIII edge (0.871 keV), making the Ne more abundant of its actual value; the Al energy is instead very different (1.52 keV vs 1.564 keV). We finally add 6 narrow absorption lines which further lower the $\chi_{\nu}^2$ ($\chi^2/dof=1178.26/631$; see Table~\ref{RGS_fit} and Figure~\ref{rgs}). We identify at least the lines of Ne IX (0.922 keV), Na X (1.12 keV) and Mg XI (1.35 keV); a line found at 1.766 keV can be instead associated to a mismodelling of Mg IX edge (1.762 keV), while for the remaining two other lines ($\sim$0.96 and 1.18 keV) the association was less certain. However, although we cannot find any other significant feature in the spectrum, 
the best-fit is still poor and the residuals are still confused, especially around the Al edge (Figure~\ref{rgs}).

\begin{figure}
\center
\includegraphics[height=6.7cm,width=8.5cm]{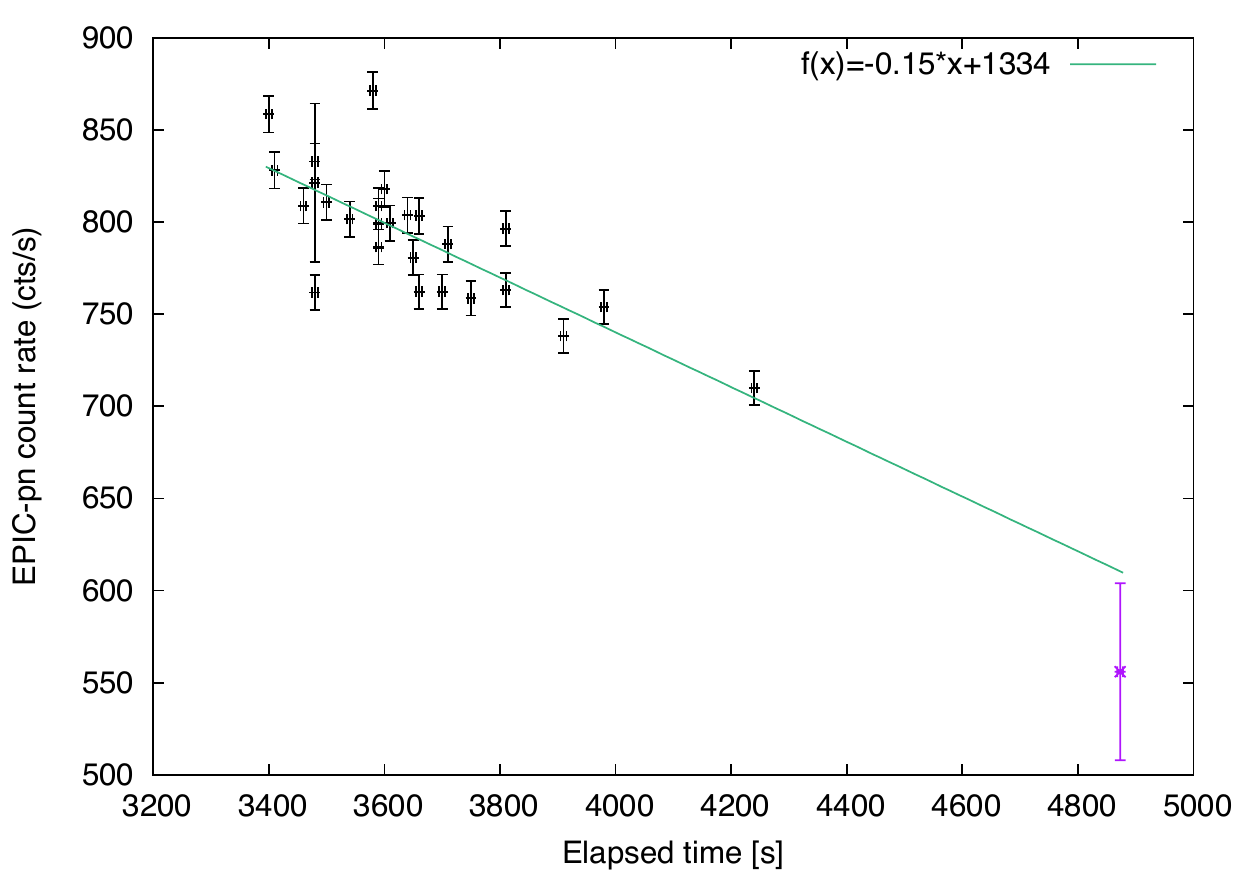}
\includegraphics[height=6.7cm,width=8.5cm]{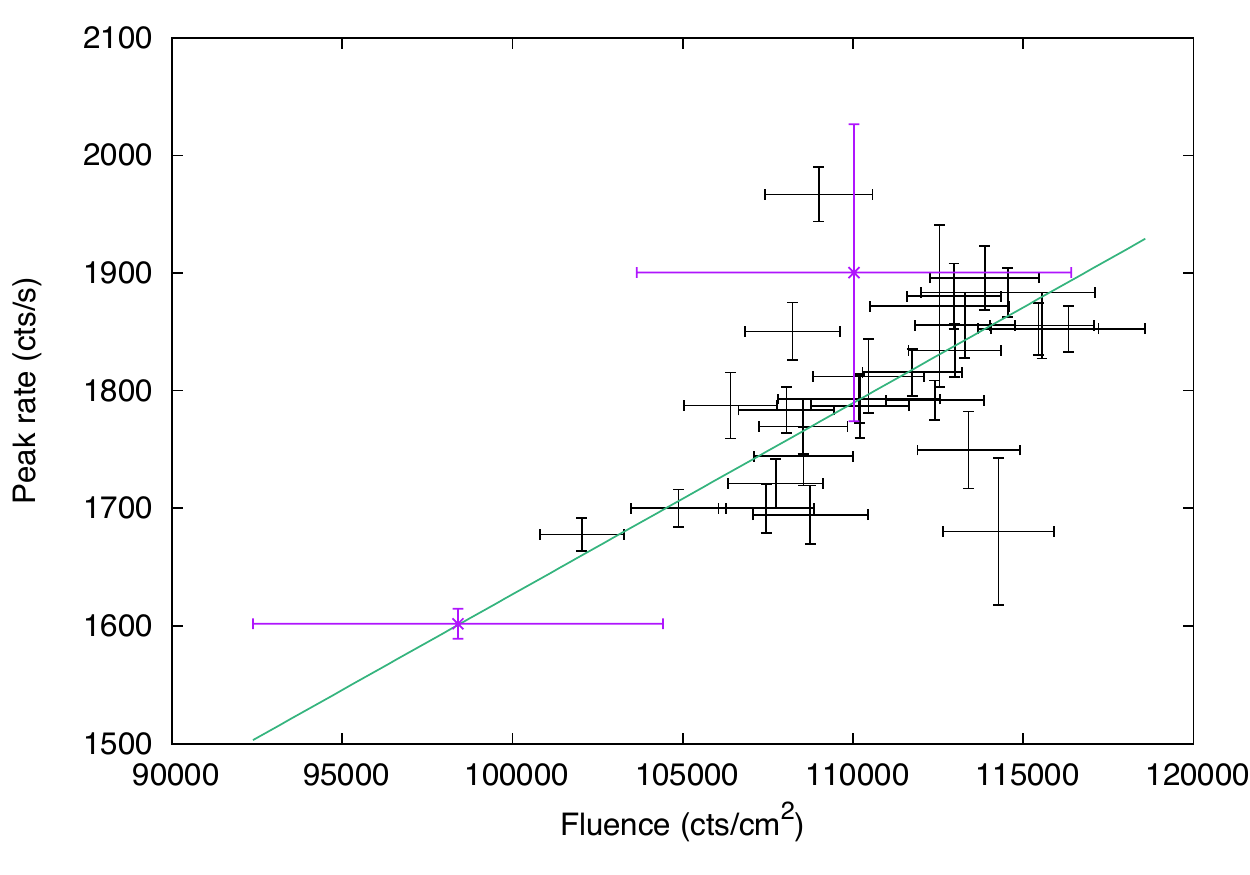}
\caption{\textit{Top}: pre-burst EPIC-pn count rate as a function of the elapsed time between a burst and the next one. A negative trend is observed. \textit{Bottom}: peak count rate versus fluence in each burst observed during the {\it XMM-Newton} observation. \newline In both plots, black points and purple crosses are bursts caught during the EPIC-pn TIMING mode and BURST mode, respectively. The green, solid lines represent the best fit of the points.}
\label{rate_vs_burst}
\end{figure}

\subsection{Burst analysis}
\label{burst_sec}

The X-ray bursts in SAX J1748.9-2021 were associated to type-I bursts because of clear correlations of the blackbody temperature and the corresponding emitting radius during their rises and decays \citep{galloway08,guver13}.
The total number of X-ray bursts observed in the EPIC-pn data (TIMING mode) is 25, with a typical duration between the rise and the return to pre-burst rate of $\sim$100 s. We model the light curve of each burst with an exponential function and we find that the typical peak count rate is $\sim$1800 cs s$^{-1}$ with an average decay folding time of $\tau=61.1\pm 0.7$ s. We note that the bursts occurred about one per hour; however, a more careful analysis shows that the elapsed time between a burst and the next presents an anti-correlation with the count rate right before the onset of the burst (Figure~\ref{rate_vs_burst}-\textit{top}). This suggests that, when the accretion rate increased, the rate of the bursts increased accordingly, as expected if the amount of matter needed to trigger the burst was more rapidly collected. In order to better investigate this process, we also calculate the fluence of each burst (i.e. the total energy released during the burst) as a function of the type-I burst peak rate. In Figure~\ref{rate_vs_burst} (\textit{bottom}), we show that the fluence increases with the burst peak rate, as expected if more material is deposited and burned on the NS surface. We note that the peak count rate does not show any saturation with the fluence, typical of bursts not reaching the Eddington limit \citep[see e.g.][]{basinska84}: this may be an effect of the {\it XMM-Newton} counting-mode because of telemetry saturation which was often triggered during the TIMING mode observation, leaving random gaps (of $\sim$8 seconds, on average) inside most of the bursts and hence possibly affecting the estimate of the total fluence. However, this effect is not present during the BURST mode light curve and we are able to estimate the correct total fluence, although with larger error bars because of the poorer statistics. In Figure~\ref{rate_vs_burst} ({\it bottom}), we show that the fluences calculated for both modes are consistent within the errors, suggesting the counting mode was not significantly contributing to the underestimation of the fluence. This result points towards the hypothesis that the type-I bursts did not reached the maximum peak luminosity (i.e. at the Eddington limit). 

\begin{figure}
\center
\includegraphics[height=6.7cm,width=8.5cm]{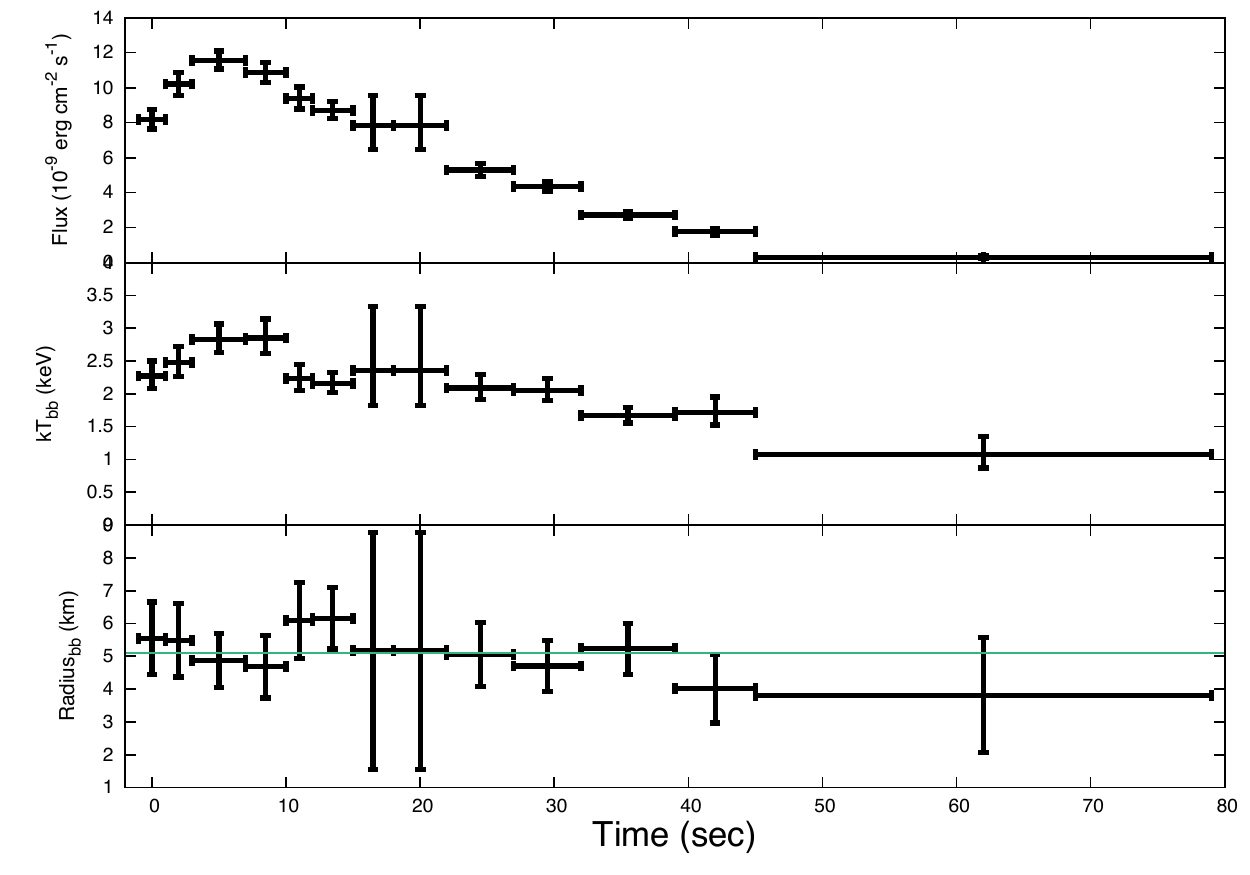}
\includegraphics[height=6.7cm,width=8.5cm]{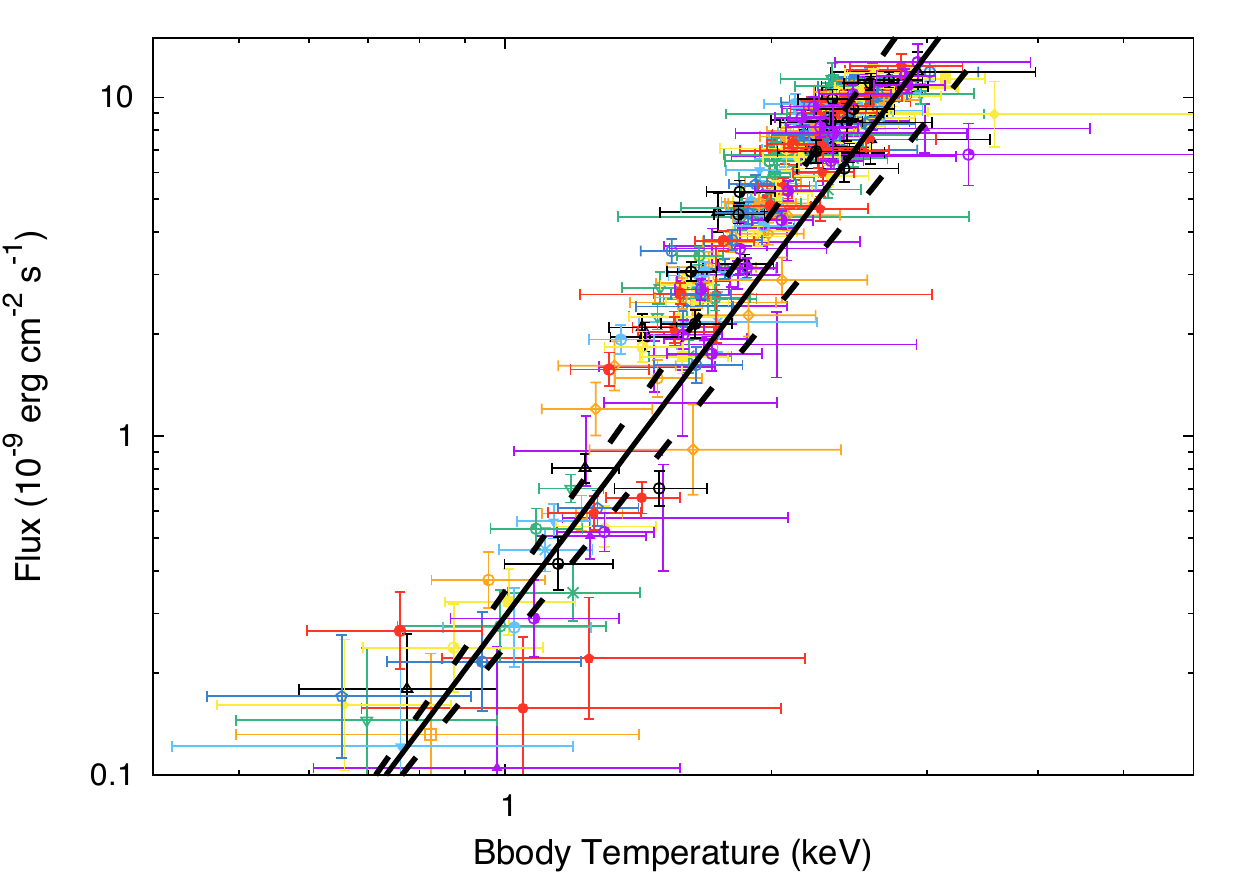}
\caption{{\it Top}: evolution of the 0.3-10.0 keV flux (\textit{upper panel}), the blackbody temperature (\textit{middle panel}) and emission radius (\textit{lower panel}) during a type-I burst. {\it Bottom}: evolution of the 0.3--10 keV flux as a function of the blackbody temperature during the bursts observed in the \textit{XMM-Newton} observation. The black, solid line represents the best fit (F$_{0.3-10keV}\propto$ T$^{3.5\pm0.1}$), while the black, dashed lines its 3 sigma upper limits.}
\label{burst9}
\end{figure}

\begin{figure}
\center
\includegraphics[height=6.7cm,width=8.5cm]{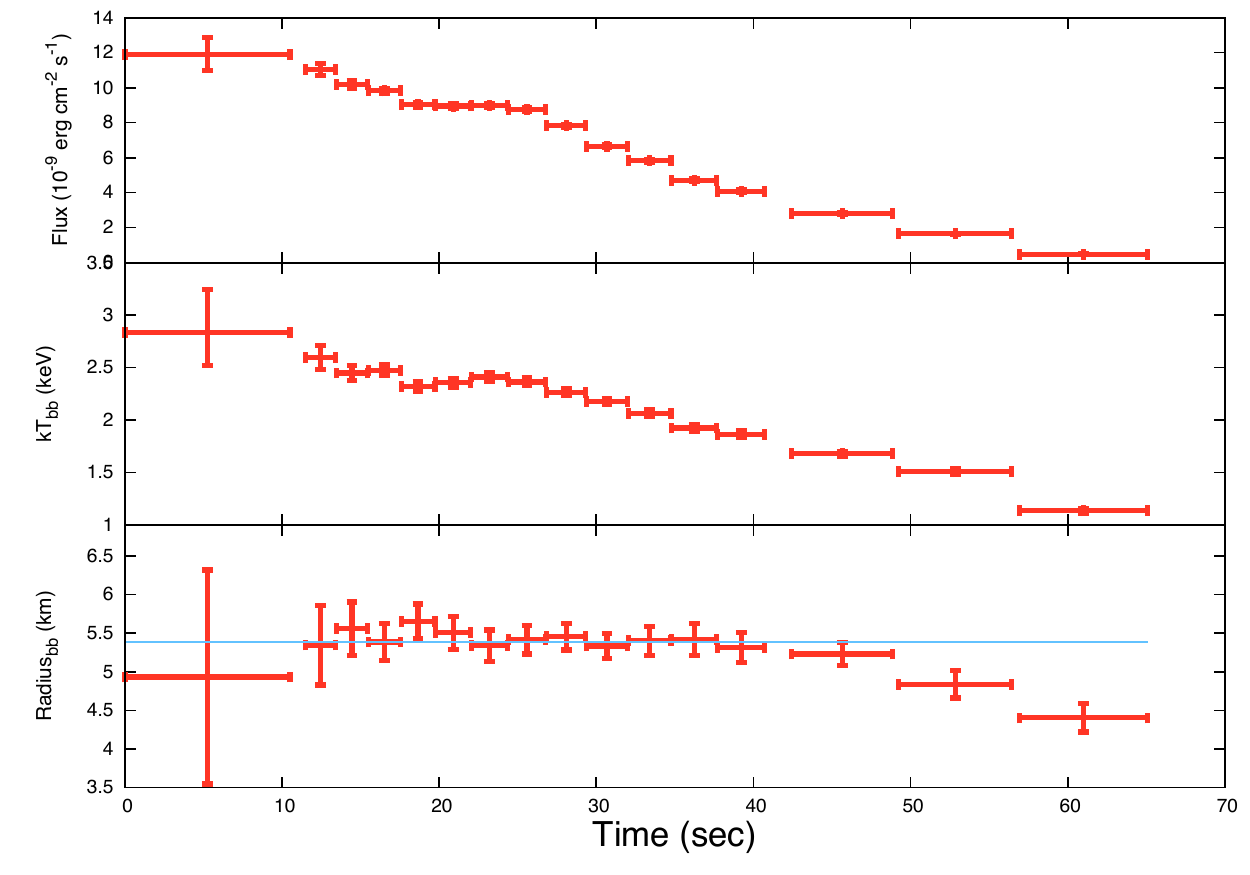}
\caption{Evolution of the 0.3-10.0 keV flux (\textit{top panel}), the blackbody temperature (\textit{center panel}) and emission radius (\textit{bottom panel}) obtained by flux-bin spectra obtained by stacking all of the type-I bursts. Beyond 40/50 s after the peak the statistics of the single spectra is very low and the spectral parameters are not reliable. The emission radius, modified by the colour correction factor ($1.35\pm0.05$) is consistent with a NS radius of 6.97--7.6 km.}
\label{burst}
\end{figure}

In order to study the spectral evolution of the type-I bursts, for each of them we create a number of EPIC-pn spectra obtained by collecting photons from small intervals (from 2s to 15s bin time according to the count rate) of their light curves. We use the averaged persistent spectrum as a background of the burst spectra, although this approach may be source of uncertainties (see e.g. \citealt{guver12_1}a,b). We note that no short-term spectral variability was observed during the {\it XMM-Newton} observation allowing us to use the averaged persistent spectrum of the whole observation. We then fit each spectrum in the 1.0--10 keV energy range with an absorbed {\sc blackbody} model. Because we do not expect significant variability of the local absorbing materials around the source, we fix the absorption neutral column density to the best-fit value of the persistent spectrum ($5.8\times10^{21}$ cm$^{-2}$). In Figure~\ref{burst9}-{\it top}, we show the flux evolution of one of the type-I bursts ({\it upper panel}) and the corresponding black body temperature ({\it middle panel}) and radius ({\it lower panel}) evolution during the rise and the decay. The absorbed peak flux is around two/three times higher than the persistent flux, and the blackbody temperature increases up to $\sim3$ keV and then it decreases down to $\sim1$ keV at the end of the decay, although we note the large error bars. The statistics of the time-binned spectra do not allow to find evidence of a clear radius expansion, and we obtain that the radius of the emitting region is almost constant at $\sim5$ km.
In addition, we show also the cooling tails of all of the X-ray bursts in Figure~\ref{burst9} ({\it bottom}), where the flux follows a relation $F\propto T_{bb}^{3.5\pm0.1}$, which is close to the expected $F\propto T_{bb}^{4}$. The discrepancy between the theoretical and observed relation can arise from a slight variation of the blackbody radius during the burst, although we do not have enough statistics to prove it. We also remark that in \citet{galloway08} and \citet{guver12_2}, the authors reported only a sub-sample of type-I bursts of SAX J1748.9-2021 showing radius expansion at a touchdown flux of $\sim4\times10^{-8}$ erg cm$^{-2}$ s$^{-1}$ in the energy range 2.5-25 keV. Extending our flux estimates to the same energy range, we evaluate an unabsorbed flux of $2.2\times10^{-8}$ erg cm$^{-2}$ s$^{-1}$, about a factor of 2 lower than the peak flux found in the \textit{RXTE} data, although the peak temperatures result compatible. Therefore, the absence of a clear saturation for the fluence and the (on average) peak flux of $2.2\times10^{-8}$ erg cm$^{-2}$ s$^{-1}$ suggest again that none of the type-I bursts observed in the \textit{XMM-Newton} observation has reached the Eddington limit.

Since the spectral properties of the bursts are widely consistent amongst each other, in order to improve the signal-to-noise ratio, we create average flux-binned spectra obtained by stacking burst epochs at the same count rate. In this way, it was possible to find a clearer decay of the blackbody temperature from $\sim$3 keV down to $\sim$1.2 keV (Figure~\ref{burst}), which is (as expected) very close to the blackbody temperature in the persistent spectrum (see Section~\ref{persistent}). In addition, where the quality statistics is rich enough, the radius appears to be rather constant and consistent with $5.39\pm0.03$ km as no radius expansions are observed. Assuming a correction factor for the NS photosphere of $1.35\pm0.05$ (e.g. \citealt{madej04, majczyna05, guver10a}; \citealt{guver10b}b), the emitting radius allows us to constrain the NS radius at 6.97--7.6 km (assuming a distance of 8.5 kpc), consistent with the values reported in \citet{guver13}.

\section{Discussion and Conclusions} 
\label{discussion}

In this work, we have analyzed in detail the spectral properties of the intermittent accreting millisecond pulsar SAX J1748.9-2021, during the brightest phase of its 2015 outburst. We made use of a high quality \textit{XMM-Newton} observation, taken along the fast rise in flux of the outburst, and the stacking of 8 days \textit{INTEGRAL} observations, taken around the peak of the outburst. 
The source showed an intense type-I X-ray burst activity ($\sim$1 burst per hour in the \textit{XMM-Newton} observation) and, because they affect the persistent emission, we carried out a separate spectral analysis for the persistent and burst emission. 
The spectra of AMXPs are generally stable during the outbursts and can be well described by a multi-components model, where the continuum is based on a Comptonization process of soft photons in a hot, optically thin electron population (kT$_e\sim30-50$ keV), possibly located close to the NS, plus a soft ($<1$ keV) component (widely accepted to be associated with the multicolour accretion disc) and, occasionally, a third thermal continuum component is also seen and likely produced by the NS surface \citep[e.g.][]{gilfanov98,gierlinski05,falanga05,patruno09b,papitto10,papitto13}.
However, \citet{in'tzand99} studied a broad-band (0.1--100 keV) {\it BeppoSAX} spectrum of SAX J1748.9-2021 during the 1998 outburst and these authors reported that a single thermal Comptonization component was able to well describe the spectral properties of the source. The characteristics of such a component were consistent with an electron population temperature of $\sim15$ keV and optical depth of $\sim3-6$, fed by soft photons of $\sim0.6$ keV. However, no additional soft components were detected.
{ Differently from the spectral state of SAX J1748.9-2021 observed by \citet{in'tzand99} (likely a hard state), we found that the {\it XMM-Newton+INTEGRAL} data presented here show that the source was likely in a soft state, with a continuum based on two soft components (a {\sc diskbb} and a {\sc bbody}), which represent the multicolour accretion disc and the NS surface respectively, a thermal Comptonization and a hard power-law tail. The thermal  Comptonization component shows that the electron temperature is quite soft (kT$_e\sim$2.1 keV) and optically thick and it is fed by seed photons whose temperature is consistent with that of the blackbody component of the NS ($\sim 1.1$ keV). }
We could roughly estimate the emission radius of the seed photons, assuming a spherical emission radius for the seed photons and that most of them are scattered in the optically thick corona. Following \citet{intZand99}, we note that the emission radius of the seed photons can be calculated as $R_{seed}=3\times10^4 d \sqrt{\frac{f_{bol}}{1+y}}/ (kT_{seed})^2$, where \textit{d} is the source distance, \textit{$f_{bol}$} is the bolometric flux of the observed Comptonizing spectrum, \textit{y} is the Compton parameter and \textit{kT$_{seed}$} is the seed photons temperature. For a Compton parameter \textit{y} of $\sim36$ (estimated by calculating the optical depth from kT$_e$ and $\Gamma_{compt}$ as shown in \citealt{zdiarski96}), an \textit{$f_{bol}$} of $\sim2.8\times10^{-9}$ erg cm$^{-2}$ s$^{-1}$, a distance of 8.5 kpc and a seed photons temperature of $\sim1.1$ keV, we estimated that the emission radius is $\sim 2\pm1$ km, { confirming a very small emitting region which may be possibly associated with a zone over the NS surface as the shock right above the accretion column. The temperature of the seed photons, similar to the NS temperature, and the size of the electron population strongly indicate that the Comptonizing medium is possibly located very close to the compact object.
This may be further confirmed by the existence of a persistent pulsations in the {\it XMM-Newton} data (Sanna et al. submitted), which point towards the existence of an hot-spot above the NS surface.}

{ We calculated that the broadband 1.0--50 keV absorbed (unabsorbed) flux is $\sim4.9\times10^{-9}$ ($\sim5.5\times10^{-9}$) erg cm$^{-2}$ s$^{-1}$, consistent with a luminosity of $L_X\sim4.2\times10^{37}$ erg s$^{-1}$ ($\sim4.7\times10^{37}$) (for a distance of 8.5 kpc), hence about 25$\%$ of the Eddington luminosity for accretion onto a 1.4 M$_{\odot}$ compact object. The {\sc diskbb, bbody, nthcomp} and {\sc power-law} components represent the $\sim$20$\%$, 25$\%$, 51$\%$ and 3$\%$ of the total unabsorbed flux, respectively.
Moreover, SAX J1748.9-2021 was quite faint at very high energies ($>10$ keV), indicating that the source was clearly in a soft state, differently from most of the AMXPs spectra. Indeed, with the exception of the three intermittent millisecond sources which have also been caught during soft states, AMXPS are usually observed in spectral states consistent with hard states. 
Furthermore, it is interesting to note that persistent pulsators among AMXPs are generally observed pulsating during hard states, while the intermittent AMPXs have been seen pulsating in both hard and soft states \citep[see e.g.][and reference therein, for a review]{patruno12}.  

In addition, some emission features are found in the broad-band spectrum of the source, in particular broad emission lines of Fe, S, Ar and Ca. The broadening of the lines may suggest that they are produced by reflection off of hard photons from the surface of the disc and smeared by the fast motion of matter in the accretion disc. Broad emission lines have been found in several AMXPs \citep{papitto09,papitto10,cackett10,wilkinson11,cackett13,papitto13} and they usually showed asymmetric profiles, which nature was associated with relativistic effects on reflected photons. From these lines, it was possible to estimate the radius at which they are produced, suggesting the existence of truncated accretion discs. Truncated discs are expected if the magnetic field is strong enough to drive the matter towards the NS magnetic polar caps. 
In our case, the Fe line ($\sim6.8$ keV) was the strongest feature in the spectrum and we tentatively fitted it with a {\sc diskline} model which can provide information about the disc region where the line was formed. We found that the estimated best-fitting inner radius of the disc is comprised between 20 and 40 R$_g$ ($\sim$40--80 km, assuming a NS of 1.4 M$_{\odot}$) with a disc inclined of 44{\textdegree} with respect to our line of sight. Also the intermittent-AMXP HETE 1900.1-2455 was characterized by an iron emission line at $\sim6.6$ keV and possibly originating at 25$\pm15$ R$_g$ \citep{papitto13}, consistent with the location of the iron line region in SAX J1748.9-2021. 
This result is quite intriguing as SAX J1748.9-2021 exhibited pulsations (persistently, as found by Sanna et al. submitted in the {\it XMM-Newton} data) during the reported soft state while the accreting millisecond pulsar HETE J1900.1-2455 in a burst occurred during an unusually soft state of the source showed only burst oscillations \citep{watt09}. However, HETE J1900.1-2455 has been also observed pulsating during hard states \citep{papitto13} unlike SAX J1748.9-2021 which, we highlight, pulsated only during soft states \citep{patruno09}. 
In particular, it has been proposed that intermittency of HETE J1900.1-2455 is due to a magnetic field generally buried under the NS surface \citep{cumming08} or, if the magnetosphere is considered, to a break-through of accreting matter into the magnetic field lines, possibly accompanied by position changes of a hot spot located close to the spin axis of the NS \citep{romanova08,lamb09}. Other hypotheses suggest limitation of the pulse amplitude related to gravitational light bending effects \citep{ozel09} or scattering of the coherent pulsation in an optically thick, hot corona around the NS \citep{titarchuk02}. The truncation of the disc found by \citet{papitto13} can be then possibly explained as an optically thin accretion flow in the regions closer to the NS, as also proposed for LMXB systems with non-pulsating NSs. However, such interpretations do not appear to be consistent with the characteristics of SAX J1748.9-2021. Therefore the two sources appear to behave in a different way and such considerations make SAX J1748.9-2021 a very peculiar source. 

In fact, our results allow us to find a coherent picture for the source. A 3$\sigma$ upper limit to the inner radius of the accretion disc would be $\sim$20 km, not very different from the 40--80 km inferred by the iron line with the {\sc diskline} model. The discrepancy between the two values may be due to very ionised inner regions of the disc, where the iron line cannot form.
This inner disc radius implies that the matter does not reach the latest stable orbit, where the magnetic field would not be able to drive the accretion towards the polar cap (i.e. giving a truncated disc) and produce the persistent pulsating emission. 
Furthermore, the existence of a possible NS emission allows us to consider that the cold and optically thick corona does not hide the compact object. This would be expected in order to explain the pulsating emission, produced more likely in the polar caps.}

We also remark the detection of a hard power-law tail, which carries $\sim$3$\%$ of the total flux. Hard tails are very often observed in soft states of LMXB spectra \citep[e.g.][]{disalvo00,disalvo01,damico01,iaria04,disalvo06,paizis06,dai07,tarana07,piraino07}. 
During this state, the hard tails do not present high energy cut-off and they provide a small percentage of the total flux emission (a few percent), although in the case of the accreting NS GX 3+1 such a component was found to carry up to $\sim$20$\%$ of the total emission \citep{pintore15}. Several scenarios were proposed in order to explain hard tails: in particular, they can be either by non-thermal Comptonization of non thermal, relativistic, electrons in a local outflow \citep[e.g.][]{disalvo00} or in a corona \citep{poutanen98}, or by the bulk motion of accreting material close to the NS \citep[e.g.][]{titarchuk98}. Another scenario suggests synchrotron emission from a relativistic jet escaping from the system \citep{markoff01} and, in the second scenario, theory expects also that radio emission would be observed \citep[e.g.][]{migliari07,homan04}, although for SAX J1748.9-2021 such a detection is still not confirmed. Future simultaneous multiwavelength observations of the source, in particular X-ray and radio, will be very useful to further investigate the nature of such a component.

Despite of the reported results and although emission lines and continuum properties are consistent, we found difficulties to find robust spectral fits when describing the broadband spectrum with a reflection continuum adopting self-consistent models (as {\sc reflionx} or {\sc rfxconv}). Hence, our results cannot strongly rule-out the possibility that the emission lines are instead broadened by Compton down-scattering due to a narrow wind shell ejected at mildly relativistic velocities at some disc radii where the local radiation force overcomes the local disc gravity \citep{laurent07} or, as another possible scenario, by Compton processes in a moderately, optically thick accretion disc corona (ADC) generated for evaporation above the accretion disc \citep{white82a,kallman89, vrtilek93}. However, in the latter case, the broadening of the line and the temperature of the electron population should be related with $\Delta \epsilon/\epsilon= (4kT_e-\epsilon)/m_ec^2$. From this relation, the electron temperature should be $\sim9.5$ keV, more than a factor of 4 higher than the temperature obtained with the {\sc nthcomp} model. Hence, we would favour an interpretation of the SAX J1748.9-2021 emission lines in terms of reflection. Assuming that the fits obtained adopting self-consistent models (although not completely stable) are reliable, we can make some considerations about the nature of the process. In particular, we could find that the ionization level $\xi$ of the reflector is consistent with $\sim1600$ erg cm s$^{-1}$; hence, from the relation $\xi=L/n R^2$ (where L is the unabsorbed X-ray luminosity of the illuminating source, $n$ is the number density of the reflecting medium and $R$ is the distance from the reflector), we estimate a distance from the reflector of $\sim50$ km, assuming a hydrogen density of $5\times10^{20}$ cm$^{-3}$ (as expected from the Shakura \& Sunyaev model for accretion around a NS, \citealt{ross07}). The distance is well consistent with that inferred from the {\sc diskline} model and, because at $\sim$50 km from the NS surface the relativistic effects are less pronounced, this can explain why we could not robustly constrain most of the parameters of the relativistic kernel ({\sc rdblur}). 

Finally, we investigated the numerous type I X-ray bursts present in the \textit{XMM-Newton} observation. We found a clear evolution of the blackbody temperature during bursts, with a peak temperature around 3 keV, although we could not find any ``touchdown'' moment because of the limited statistics. The relation between fluence and peak count rate suggests that none of the type-I X-ray bursts was able to reach the Eddington limit.
However, it was possible to estimate the radius which is responsible for the emission of such a blackbody component, inferring a value of 7.0-7.6 km (assuming a distance of 8.5 kpc). This value is consistent within the error of the previous estimates ($8.18\pm1.6$ km or $10.93\pm2.09$ km; \citealt{guver13}) obtained by analyzing RXTE data. The average unabsorbed 0.3--10 keV peak flux is $\sim$1.2$\times10^{-8}$ erg cm$^{-2}$ s$^{-1}$, corresponding to a 0.3--10 keV luminosity of $1\times10^{38}$ erg s$^{-1}$ (assuming a distance of 8.5 kpc). This value indicates that the burst reached a peak luminosity of $\sim$50$\%$ of the Eddington luminosity for a 1.4 M$_{\odot}$ NS.

\section*{Acknowledgements} 

We thank N. Schartel, who made this ToO observation possible in the Director Discretionary Time, and the {\it XMM-Newton} team who performed and supported this observation. We also thank the unknown referee for his/her help in improving the paper.
Authors acknowledge financial contribution from the agreement ASI-INAF I/037/12/0. We gratefully acknowledge the Sardinia Regional Government for the financial support (P. O. R. Sardegna F.S.E. Operational Programme of the Autonomous Region of Sardinia, European Social Fund 2007-2013 - Axis IV Human Resources, Objective l.3, Line of Activity l.3.1). This work was partially supported by the Regione Autonoma della Sardegna through POR-FSE Sardegna 2007-2013, L.R. 7/2007, Progetti di Ricerca di Base e Orientata, Project N. CRP-60529. The High-Energy Astrophysics Group of Palermo acknowledges support from the Fondo Finalizzato alla Ricerca (FFR) 2012/13, project N. 2012-ATE-0390, founded by the University of Palermo. MDS thanks the Dipartimento di Fisica e Chimica, Universita' di Palermo, for the hospitality.

\addcontentsline{toc}{section}{Bibliography}
\bibliographystyle{mn2e}
\bibliography{biblio}

\end{document}